\documentclass[
     final            
  ]
  {aipproc}

\layoutstyle{8x11single}

\begin{document}

\title{Dynamical coupled-channels study of meson production reactions from EBAC@JLab}


\author{Hiroyuki~Kamano}{
  address={Excited Baryon Analysis Center (EBAC), Thomas Jefferson National Accelerator Facility, Newport News, VA 23606, USA}
}

\begin{abstract}
We present the current status of a combined and simultaneous analysis 
of meson production reactions based on a dynamical coupled-channels (DCC) model,
which is conducted at Excited Baryon Analysis Center (EBAC) of Jefferson Lab.
\end{abstract}

\maketitle


\section{Introduction}
An understanding of the spectrum and structure of the excited nucleon ($N^*$) states 
is a fundamental challenge in the hadron physics. 
The $N^*$ states, however, couple strongly to the meson-baryon continuum states and 
appear only as resonance states in $\pi N$ and $\gamma N$ reactions. 
Such strong couplings to the meson-baryon continuum states influence significantly 
the $N^*$ properties and cannot be neglected in extracting the $N^*$ parameters from 
the data and giving physical interpretations. 
It should also be emphasized that at present even the existence is still uncertain for most of
the $N^\ast$ states higher than 1.6 GeV~\cite{said06}.
This will be because most of the $N^\ast$ information were extracted 
only from the $\pi N\to \pi N$ and $\gamma N\to \pi N$ analysis.
It is thus well recognized nowadays that the comprehensive study of 
\emph{all} relevant meson production reactions with 
$\pi N$, $\eta N$, $\pi\pi N$, $KY$, $\omega N$, $\cdots$ final states 
based on a coupled-channels framework
is inevitable for reliable extraction of such higher $N^\ast$ states.

To make a progress to this direction, the Excited Baryon Analysis Center (EBAC) of Jefferson Lab
is conducting a dynamical coupled-channels (DCC) analysis
of $\pi N$, $\gamma N$, and $N(e,e')$ reactions in the resonance region.
The analysis is based on an unitarized coupled-channels model,
the EBAC-DCC model (see Ref.~\cite{msl} for the details),
within which the couplings among relevant meson-baryon reaction channels 
are fully taken into account.

In this model, the reaction amplitudes $T_{\alpha,\beta}(p,p';E)$ are calculated from
the following coupled-channels integral equations,
\begin{equation}
T_{\alpha,\beta}(p,p';E)= V_{\alpha,\beta}(p,p') + \sum_{\gamma}
\int_0^{\infty} q^2 d q  V_{\alpha,\gamma}(p, q)
G_{\gamma}(q ,E)
T_{\gamma,\beta}( q  ,p',E) \,, \label{eq:cct}\\
\end{equation}
\begin{equation}
V_{\alpha,\beta}= v_{\alpha,\beta}+
\sum_{N^*}\frac{\Gamma^{\dagger}_{N^*,\alpha} \Gamma_{N^*,\beta}}
{E-M^*} \, ,
\label{eq:ccv}
\end{equation}
where $\alpha,\beta,\gamma = \gamma N, \pi N, \eta N, \pi\pi N, K\Lambda, K\Sigma,\omega N$
(the $\pi\pi N$ channel contains the quasi two-body $\pi \Delta, \rho N, \sigma N$ channels);
$v_{\alpha,\beta}$ is a meson-exchange interaction including only ground state mesons and 
baryons, which is derived from phenomenological Lagrangians; 
$\Gamma_{N^*,\beta}$ describes the excitation of the nucleon to a bare $N^*$ state 
with a mass $M^*$; 
$G_{\gamma}(q ,E)$ is the meson-baryon Green function for the channel $\gamma$.
The second term of Eq.~(\ref{eq:ccv}) thus describes the $s$-channel exchange
of bare $N^\ast$ states.
Through the reaction processes, the bare $N^\ast$ states couple to the meson-baryon
continuum states (reactions channels) and become resonance states.
On the other hand, the meson-exchange potential $v_{\alpha,\beta}$ can also 
generate molecule-like resonances dynamically.
The physical nucleon resonances will be a ``superposition'' of 
these two pictures in general.

During the developing stage of EBAC in 2006-2009,  
hadronic and electromagnetic parameters of the EBAC-DCC model were determined by analyzing 
$\pi N\rightarrow \pi N$~\cite{jlms07} and $\pi N \rightarrow \eta N$~\cite{djlss08}
up to $W=2$ GeV, and $\gamma N \rightarrow \pi N$~\cite{jlmss08} and $N(e,e'\pi)N$~\cite{jklmss09}
up to $W=1.6$ GeV. Then, the model was applied to $\pi N\to \pi\pi N$~\cite{kjlms09} and
$\gamma N\to\pi\pi N$~\cite{kjlms09-2} to predict cross sections and examine consistency of 
the coupled-channels framework.
Also, nucleon resonance poles were extracted from the model and a new interpretation for 
the dynamical origin of $P_{11}$ nucleon resonances was proposed~\cite{sjklms10}.

Although the model constructed in the early analysis successfully describes the data 
in the wide energy region, it is still far from our goal because
only $\pi N$ and $\eta N$ channels are taken into account in the fit
and each reaction is analyzed rather separately
(e.g., in the analysis of $\gamma N\to \pi N$,
the hadronic parameters fitted to the $\pi N$ scattering data are used and,
only the electromagnetic parameters are varied).
To proceed further, we have started a full combined analysis of
$\pi N$, $\eta N$, $\pi\pi N$, $KY$, $\omega N$ channels recently, in which
all parameters are varied simultaneously.
We are currently performing a combined analysis of 
$\pi N, \gamma N \rightarrow \pi N, \eta N, K\Lambda, K\Sigma$ reactions as a first step.
In this contribution, we present the status of our current effort for the analysis.
\section{Current status of the EBAC-DCC analysis}
\begin{figure}[tb]
\includegraphics[width=0.9\textwidth,clip]{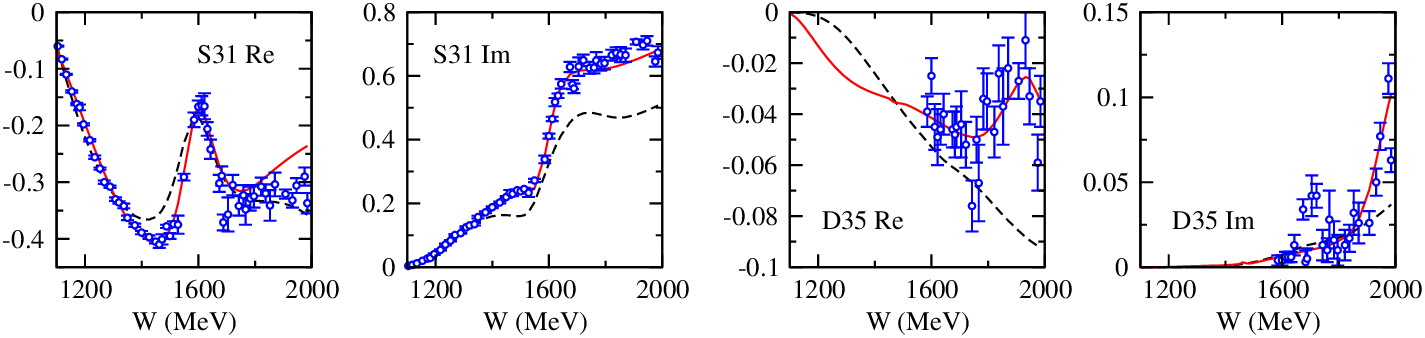}
\caption{
The partial wave amplitudes of the $\pi N$ scattering.
(Red solid curves) The current result (\textit{preliminary}). 
(Black dashed curves) Our previous model~\cite{jlms07}. 
The data points are the energy-independent solution of SAID, 
which are taken from Ref.~\cite{saidweb}.
}
\label{fig:pin}
\end{figure}

\underline{The $\pi N\to \pi N$ and $\gamma N \to \pi N$ reactions}:
The partial wave amplitudes of the $\pi N$ scattering are shown in Fig.~\ref{fig:pin}.
Here we present only the result of two partial waves, $S_{31}$ and $D_{35}$, where
a significant improvement has been achieved from our early model~\cite{jlms07}.
Other partial waves are the same quality as the early model.
The improvement in the $D35$ partial wave is mainly due to an
inclusion of one bare $N^\ast$ state.

\begin{figure}[tb]
\includegraphics[width=0.43\textwidth,clip]{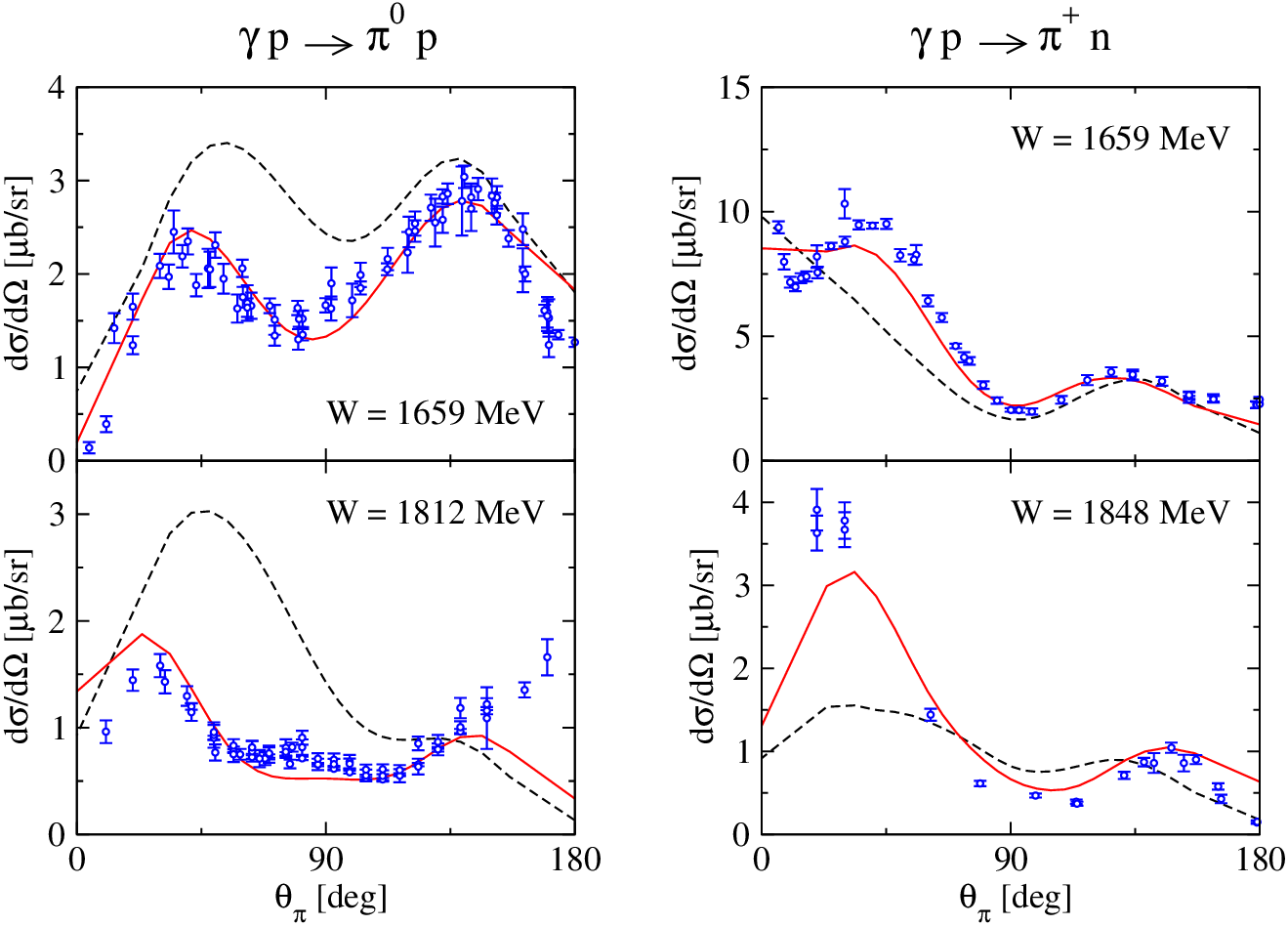}~~~
\includegraphics[width=0.43\textwidth,clip]{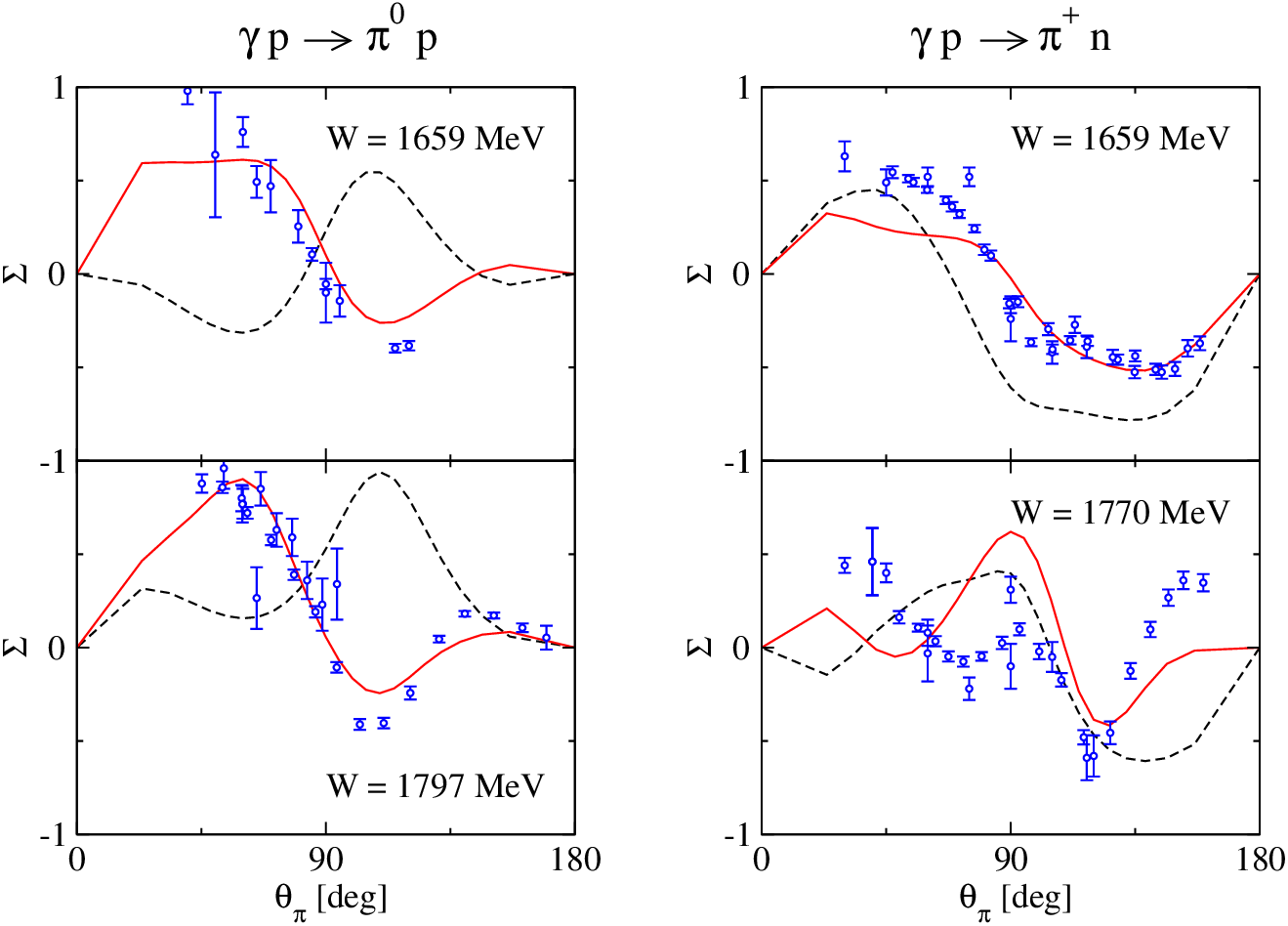}
\caption{
The differential cross sections (left two columns)
and the photon asymmetry (right two columns) of the $\gamma p \to \pi^0 p, \pi^+ n$ reactions 
at high energies $W > 1.6$ GeV.
(Red solid curves) The current result (\textit{preliminary}). 
(Black dashed curves) Our previous model~\cite{jlmss08}. 
The data are taken from the database of Ref.~\cite{saidweb}.
}
\label{fig:gpin}
\end{figure}

The current $\gamma N\to \pi N$ analysis has two major progresses
over our previous analysis constructed in Ref.~\cite{jlmss08}: 
(a) the energy region for the analysis is extended from 
$W\leq1.6$ GeV to $W\leq2$ GeV, and (b) data of the unpolarized differential cross 
section and all measured polarization observables are taken into account in the fit, 
while only differential cross sections and photon asymmetry $\Sigma$ are used 
in our previous analysis.
In Fig.~\ref{fig:gpin}, our preliminary result (red solid curves) of $\gamma N\to \pi N$
at the energies $W> 1.6$ GeV 
is compared to our previous model (black dashed curves) as well as the experimental data.
We can actually observe a visible improvement of the model in $W>1.6$ GeV.
\\

\begin{figure}[tb]
\includegraphics[width=0.9\textwidth,clip]{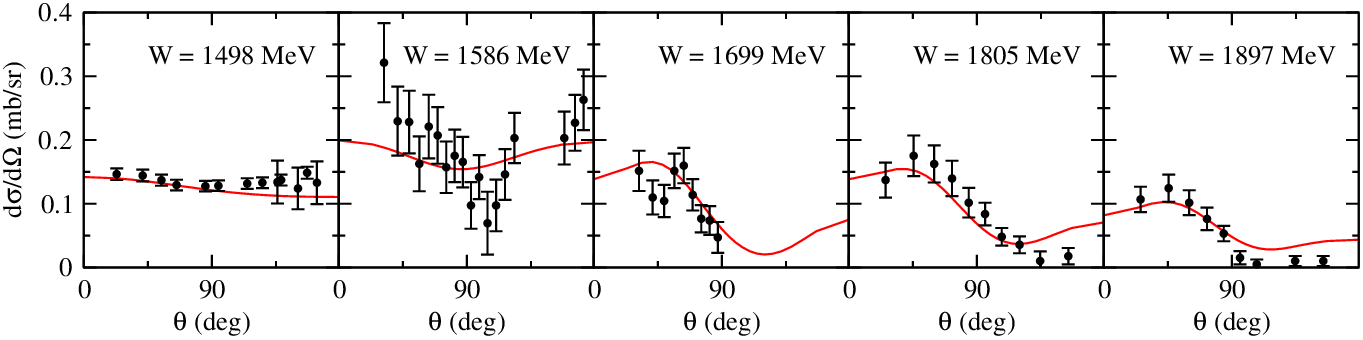}
\caption{
The differential cross sections of the $\pi^- p \to \eta n$ reactions.
Red solid curves are the current result (\textit{preliminary}). 
See Ref.~\cite{djlss08} for the details of the data used.
}
\label{fig:pen}
\end{figure}

\underline{The $\pi N\to \eta N$ reaction}:
It is known that for the $\pi^- p\to \eta n$ reaction
underlying inconsistencies exist among data sets from different experimental
groups~\cite{djlss08}.
In our analysis, we follow the strategy described in Ref.~\cite{djlss08} for 
how to select the data to be used for our analysis.
In Fig.~\ref{fig:pen}, we present the current status of the $\pi^- p \to \eta n$ reactions.
We find that at present our result describes reasonably well the differential cross section 
data up to $W=1.9$ GeV.
The analysis of $\gamma p\to \eta p$ is also underway. 
\\

\begin{figure}[tb]
\includegraphics[width=0.7\textwidth,clip]{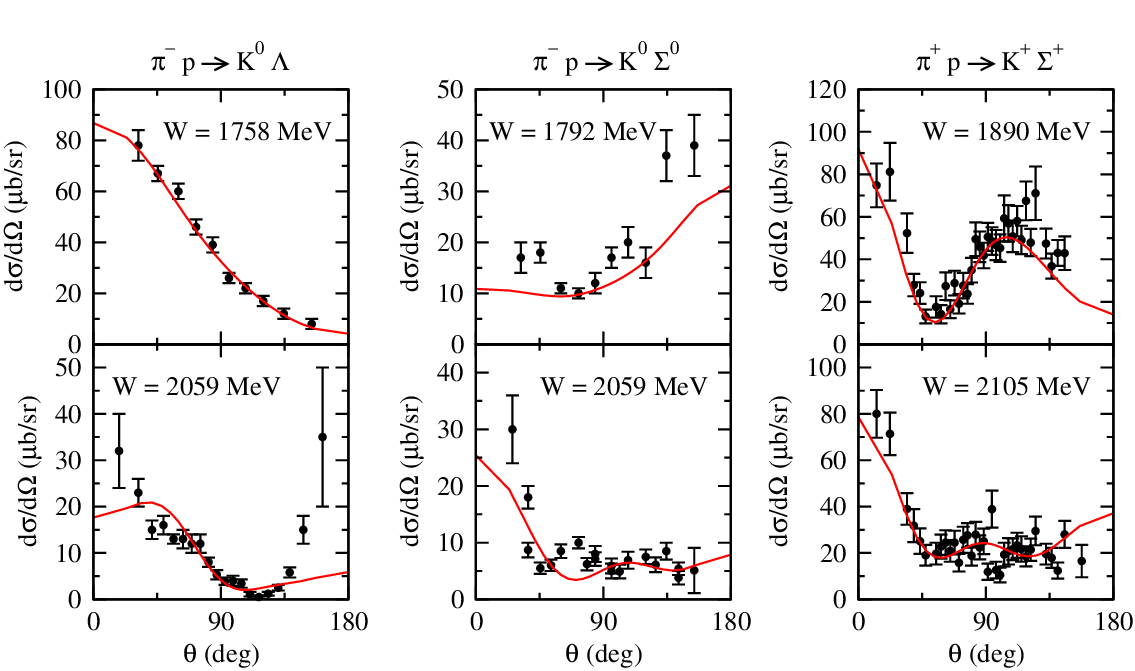}
\caption{
The preliminary result for the differential cross sections of 
$\pi^+ p \to  K^+ \Sigma^+$ (left panels),
$\pi^- p \to  K^0 \Sigma^0$ (middle panels), and
$\pi^- p \to  K^0 \Lambda^0$ (right panels).
The data are from Refs.~\cite{brunoky,ksigma}.
}
\label{fig:pkl}
\end{figure}

\underline{The $\pi N\to KY$ and $\gamma N\to KY$ reactions}:
Now we move to showing the current status of the analysis for the $KY$ reactions.
In Fig.~\ref{fig:pkl}, we present the differential cross sections of
the $\pi N\to KY$ reactions for three different charge states.
At present we have included the data up to $W=2.1$ GeV for the analysis.
The current result describes the data of the considered energy region reasonably well.

Finally, we present the current status of the $\gamma p\to K^+\Lambda$ analysis.
This strangeness-production reaction is expected to be one of the most promising reactions
to provide critical information for confirming/rejecting not well-established $N^\ast$s
and/or discovering new $N^\ast$s.
Because of this, measurement of the polarization observables, which are more exclusive 
than unpolarized cross section, for 
$\gamma p\to K^+\Lambda$ becomes very active at electron beam facilities. 
For example, first measurement of the $O_{x'}$, $O_{z'}$, and $T$ asymmetries 
has been reported recently by GRAAL~\cite{graal09}.
Furthermore, the so-called ``(over-) complete experiments'' is planned at CLAS,
in which \emph{all} 15 polarization observables are measured, and so
extensive data will be available in near future.

\begin{figure}[tb]
\includegraphics[width=0.9\textwidth,clip]{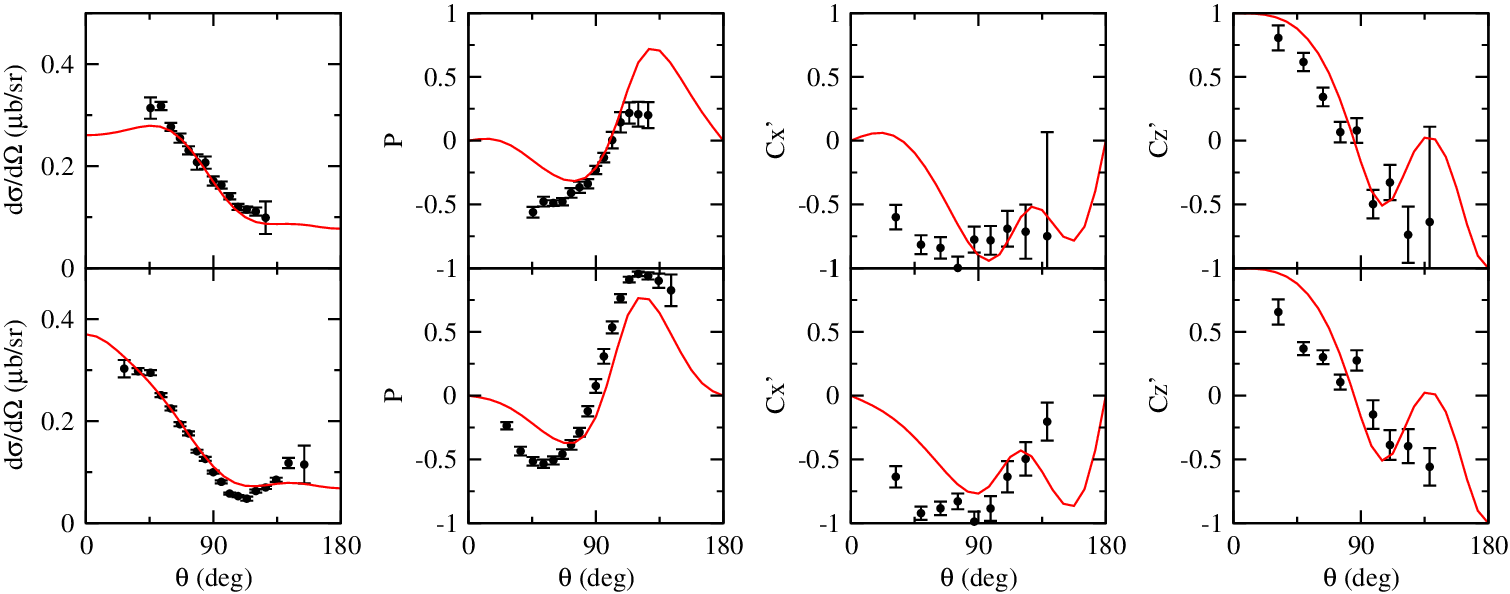}
\caption{
The preliminary result for the differential cross sections and $P$, $C_{x'}$, and $C_{z'}$
asymmetries of $\gamma p \to K^+\Lambda$.
The upper (lower) panels are the result at $E_\gamma = 1250$ MeV
($E_\gamma = 1650$ MeV).
The data are from Refs.~\cite{g11a10,g1c07}.
}
\label{fig:gkl}
\end{figure}

In Fig.~\ref{fig:gkl}, the differential cross sections and the polarization observables
$P, C_{x'}, C_{z'}$ are compared with those measured in the CLAS-g11a~\cite{g11a10} 
and CLAS-g1c~\cite{g1c07} experiments.
Here we note that care must be taken in calculating the polarization observables
because incompatibility exist in the expressions of the observables 
in the literature~(see Ref.~\cite{shkl} for the detail).
In our analysis, we follow the definitions explicitly described in Ref.~\cite{shkl}.
Although it must be further improved, our model describes qualitatively
the available differential cross sections and polarization observables.
In order to describe the data of all polarization observables measured in upcoming 
experiments at CLAS, however, we may need to introduce additional bare $N^\ast$ states 
to reproduce such extensive data.
Then the introduced bare $N^\ast$ states may generate new resonance states.
\\

In summary, the Excited Baryon Analysis Center of Jefferson Lab makes a continuous 
effort for a combined and simultaneous coupled-channels analysis of all relevant meson-production 
reactions toward the ultimate goal of establishing the $N^\ast$ spectrum and extracting 
$N^\ast$ parameters.
We are currently performing a combined analysis of 
the $\pi N,\gamma N\to \pi N,\eta N, K\Lambda, K\Sigma$ reactions.
Although further improvements are necessary, our current model describes the reactions
reasonably well from threshold up to $W\sim 2$ GeV.
Once this analysis is completed, we will gradually extend our analysis by including other 
reactions with final states such as $\pi \pi N$ and $\omega N$.


\begin{theacknowledgments}
The author would like to thank B.~Juli\'a-D\'{\i}az, T.-S.~H.~Lee, 
A.~Matsuyama, S.~X.~Nakamura, T.~Sato, and N.~Suzuki for their collaborations at EBAC.
This work was supported by 
the U.S. Department of Energy, Office of Nuclear Physics Division, under 
Contract No. DE-AC05-06OR23177 
under which Jefferson Science Associates operates the Jefferson Lab.
\end{theacknowledgments}



\bibliographystyle{aipproc}   

\begin{thebibliography}{9}

\bibitem{said06}
R.~A.~Arndt, W.~J.~Briscoe, I.~I.~Strakovsky, and R.~L.~Workman,
\emph{Phys. Rev. C} \textbf{74}, 045205 (2006).

\bibitem{msl}
A. Matsuyama, T. Sato, and T.-S. H. Lee, 
\emph{Phys. Rep.} {\bf 439}, 193 (2007).

\bibitem{jlms07}
B.~Juli\'a-D\'{\i}az, T.-S.~H.~Lee, A.~Matsuyama, and T.~Sato,
\emph{Phys. Rev. C} \textbf{76}, 065201 (2007).

\bibitem{djlss08}
J.~Durand, B.~Juli\'a-D\'{\i}az, T.-S.~H.~Lee, B.~Saghai, and T.~Sato,
\emph{Phys. Rev. C} \textbf{78}, 025204 (2008).

\bibitem{jlmss08}
B.~Juli\'a-D\'{\i}az, T.-S.~H.~Lee, A.~Matsuyama, and T.~Sato,and L.~C.~Smith,
\emph{Phys. Rev. C} \textbf{77}, 045205 (2008).

\bibitem{jklmss09}
B.~Juli\'a-D\'{\i}az, H. Kamano, T.-S.~H.~Lee, A.~Matsuyama, T.~Sato, N. Suzuki, 
Phys.\ Rev.\ C {\bf 80}, 025207 (2009).  

\bibitem{kjlms09}
H.~Kamano, B.~Juli\'a-D\'{\i}az, T.-S.~H.~Lee, A.~Matsuyama, and T.~Sato,
Phys.\ Rev.\ C {\bf 79}, 025206 (2009).

\bibitem{kjlms09-2}
H.~Kamano, B.~Juli\'a-D\'{\i}az, T.-S.~H.~Lee, A.~Matsuyama, and T.~Sato,
Phys.\ Rev.\ C {\bf 80}, 065203 (2009).

\bibitem{sjklms10}
N. Suzuki, B. Julia-Diaz, H. Kamano, T.-S. H. Lee, A. Matsuyama, T. Sato,
Phys. Rev. Lett. {\bf 104}, 042302 (2010).

\bibitem{saidweb}
CNS Data Analysis Center, GWU, http://gwdac.phys.gwu.edu .

\bibitem{brunoky}
B.~Juli\'a-D\'{\i}az, B.~Saghai, T.-S.~H.~Lee, F.~Tabakin,
Phys. Rev. C {\bf 73}, 055204 (2006), references therein.

\bibitem{ksigma}
D.~J.~Candlin \textit{et al.},
Nucl. Phys. {\bf B226}, 1 (1983).

\bibitem{shkl}
A.~Sandorfi, S.~Hoblit, H.~Kamano, T.-S.~H. Lee,
arXiv:0912.3505 [nucl-th]; in preparation.

\bibitem{g11a10}
M.~E.~McCracken \textit{et al.} (CLAS collaboration),
Phys. Rev. C {\bf 81}, 025201 (2010).

\bibitem{g1c07}
R.~K.~Bradford \textit{et al.} (CLAS collaboration),
Phys. Rev. C {\bf 75}, 035205 (2007).

\bibitem{graal09}
A.~Lleres \textit{et al.} (GRAAL collaboration),
Eur. Phys. J. A {\bf 39}, 149 (2009).

\end{thebibliography}

\end{document}